\documentclass{article}
\pdfoutput=1
\usepackage[utf8]{inputenc}
\usepackage{amsmath}
\usepackage{indentfirst}
\usepackage{amssymb}
\usepackage{graphicx}
\graphicspath{ {C:/Users/hp/Documents/math publication/Def Publication/} }

\title{Sustainability of collusion and market transparency in a sequential search market: a generalization}
\author{Jacopo De Tullio*, Giuseppe Puleio**
\\ * Department of Decision Sciences, Bocconi University, Italy
\\ ** Department of Finance, Bocconi University, Italy}
\date{May 2021}

\begin{document}

\maketitle
\begin{abstract}
	The present work generalizes the analytical results of Petrikaite (2016) to a market where more than two firms interact. As a consequence, for a generic number of firms in the oligopoly model described by Janssen et al (2005), the relationship between the critical discount factor which sustains the monopoly collusive allocation and the share of perfectly informed buyers is non-monotonic, reaching a unique internal point of minimum. The first section locates the work within the proper economic framework. The second section hosts the analytical computations and the mathematical reasoning needed to derive the desired generalization, which mainly relies on the Leibniz rule for the differentiation under the integral sign and the Bounded Convergence Theorem.   
\end{abstract} 
\section{Introduction}

The work of Petrikaite (2016) analyses the sustainability of the monopoly collusive allocation as a \textit{Subgame Perfect Equilibrium} (SPE) of the super-game extension of the model of costly sequential search described by Janssen et al (2005)\footnote{We refer to the case of full consumer participation addressed by Janssen et al (2005). Within this framework, this model should be thought of as the inelastic demand version of the seminal model of Stahl (1989).} in the case of a duopoly. 
A well established result of the theory of industrial organization asserts that in a infinitely repeated oligopoly game with almost perfect monitoring the monopoly collusive allocation constitutes a SPE provided that the inter-temporal discount factor $\delta \in (0,1)$ is not inferior than a certain critical value.\footnote{That is, the factor through which firms, at time $t$, discount expected profits which will mature at time $t+1$. Indeed, if firms are sufficiently patient, they are willing to renounce to the short term gains that they could achieve best-responding against the collusive agreement, in order to preserve the future \textit{supracompetitive} margins that collusion is expected to bring about. This is the classic model of tacit collusion when firms play grim-trigger strategies. See, \textit{inter alia}, Motta (2004) and Tirole (1988).} The minimum discount factor $\delta^{*}$ is related to the one-shot Nash Equilibrium profits $\pi^{*}$, the collusive profits $\pi^{c}$ and the deviation profits $\pi^{d}$ by the following equation:
$$ \delta^{*}=\frac{\pi^{d}-\pi^{c}}{\pi^{d}-\pi^{*}} \in (0,1) $$
\par
 	    In the model of Janssen et al (2005), consumers search sequentially in a homogeneous market with perfect recall; a portion $\lambda$ of them can become aware of all the different quotations listed by the firms at no cost, while the remaining part of buyers suffer a constant search cost $s$, smaller than their maximum willingness to pay $v$, for each price they might desire to observe, except for the first one. Demand is inelastic: each consumer purchases one unit of output, as long as the minimum price of which she is aware does not exceed her maximum willingness to pay. Given this environment, in a monopoly collusive allocation it holds that $\pi^{c}=\frac{v}{N}$, $\pi^{d}=v(\frac{1-\lambda}{N}+\lambda)$ and $\pi^{*}=\frac{p^{*}(1-\lambda)}{N}$ where $N$ represents the number of firms in the market and $$p^\ast=\frac{s}{1-\int_{0}^{1}\frac{dy}{1+\frac{\lambda}{1-\lambda}Ny^{N-1}}}=\frac{s}{1-G(\lambda;N)}$$
represents the endogenous reservation price, which is defined as the price that makes imperfectly informed buyers indifferent between purchasing at the price at hand or carrying out a further search through the market. $N$, $\lambda$, $v$ and $s$ are all exogenous, and it is assumed that $0<s<v$ and $1>\lambda\geq{\hat{\lambda}}$, so that a reservation price equilibrium is achieved\footnote{A reservation price equilibrium exists if and only if $\lambda\geq{\hat{\lambda}}$ where $\hat{\lambda}
$ satisfies $1-\int_{0}^{1}\frac{dy}{1+\frac{\hat{\lambda}}{1-\hat{\lambda}}Ny^{N-1}}=\frac{s}{v}$, given $s \in (0,v)$. The reservation price will correspond to the upper bound of the prices distribution; see Pennerstorfer et al (2016) and Janssen et al (2005) for more details. The case of a no-reservation price equilibrium is significantly less interesting, since in that case the critical discount factor will depend only on the number of firms in the market, being insensitive to variations in the relative number of perfectly informed buyers. }, while $y \in [0,1]$ is an auxiliary variable derived from the equilibrium distribution of prices. 
\\
In the light of what has been specified above, it is possible to write: 
\begin{equation}
\delta^\ast=\frac{\lambda\left(N-1\right)}{1+\lambda\left(N-1\right)-\frac{p^\ast\left(1-\lambda\right)}{v}}
\end{equation} and 
\begin{equation*}
\begin{split}
\frac{\partial{\delta^\ast}}{\partial\lambda}=\frac{1}{\left(\pi^d-\pi^\ast\right)^2}\left[\frac{\partial\pi^\ast}{\partial\lambda}\left(\pi^d-\pi^c\right)+\frac{\partial\pi^d}{\partial\lambda}\left(\pi^c-\pi^\ast\right)\right]=\\
=\frac{1}{\left(\pi^d-\pi^\ast\right)^2}\left[\frac{v}{N^2}\left(N-1\right)\left(v-p^\ast+\frac{\partial p^\ast}{\partial\lambda}\left(1-\lambda\right)\lambda\right)\right]
\end{split}
\end{equation*}
Let $$\mathrm{\Gamma}\left(\lambda;N\right)=v-p^\ast+\frac{\partial p^\ast}{\partial\lambda}\left(1-\lambda\right)\lambda$$
\\
Petrikaite (2016) shows that $\frac{\partial\Gamma\left(\lambda;2\right)}{\partial\lambda}|_{\ \lambda\ \in\left(0,1\right)} >0$; \textit{a fortiori}, $\frac{\partial\Gamma\left(\lambda;2\right)}{\partial\lambda}|_{\lambda\ \in\left(\hat{\lambda},1\right)}>0$. 
As $\lim_{\lambda\to\hat{\lambda}}{\mathrm{\Gamma}\left(\lambda;2\right)}= v^2\left(\frac{1}{v}-\frac{2\hat{\lambda}}{\left(1+\hat{\lambda}\right)s}\right)<0$ and $\lim_{\lambda\to1}{\mathrm{\Gamma}\left(\lambda;2\right)}= v-s > 0$, it is possible to deduce that, given $N=2$ and $s \in (0,v)$, $\exists\ \widetilde{\lambda}$ such that $\forall\ \lambda\in\ (\hat{\lambda},\widetilde{\lambda}]$ $\frac{\ \partial\delta^\ast}{\partial\lambda}\ \leq0$ and $\forall\ \lambda\in\ \left(\widetilde{\lambda},1\right)$ $\frac{\ \partial\delta^\ast}{\partial\lambda}>0$.
\\
\\

\section{The general case}
	The aim of this section is to show that the results which apply to a duopoly can be extended to a case where a higher number of firms interact: provided that the share of shoppers is sufficiently high so that a reservation price equilibrium is achieved, the critical discount factor will be first decreasing and later increasing in this parameter, reaching a unique internal point of minimum. Hence, in what follows, $N>2$ ($N\in\mathbb{N}$)\footnote{Clearly, from an economic point of view, it is reasonable to restrict ourselves to $N\leq10$.}.
\subsection{Deriving the inequality}
Consider again 
\begin{equation}
\mathrm{\Gamma}\left(\lambda;N\right)=v-p^\ast+\frac{\partial p^\ast}{\partial\lambda}\left(1-\lambda\right)\lambda=
\end{equation}
$$=v-p^\ast-\left(1-\lambda\right)\lambda\left(\frac{p^\ast}{1-G\left(\lambda;N\right)}\cdot\ \int_{0}^{1}{\frac{Ny^{N-1}}{\left(\left(1-\lambda\right)+\lambda N y^{N-1}\right)^2}dy}\right)$$
Notice that $$\int_{0}^{1}{\frac{Ny^{N-1}}{\left(\left(1-\lambda\right)+\lambda N y^{N-1}\right)^2}dy}=\frac{1}{\left(1-\lambda\right)^2}\int_{0}^{1}{\frac{Ny^{N-2}y}{\left(1+\frac{\lambda}{\left(1-\lambda\right)}Ny^{N-1}\right)^2}dy=}$$ $$=\frac{1}{\lambda\left(N-1\right)}\left[\frac{G\left(\lambda;N\right)\left(1+\lambda\left(N-1\right)\right)-\left(1-\lambda\right)}{\left(1+\lambda\left(N-1\right)\right)\left(1-\lambda\right)}\right]>0$$
So, equation (2) can be rewritten as $$\mathrm{\Gamma}\left(\lambda;N\right)=v-p^\ast\left\{1+\left(\frac{G\left(\lambda;N\right)\left(1+\lambda\left(N-1\right)\right)-\left(1-\lambda\right)}{\left(N-1\right)\left(1+\lambda\left(N-1\right)\right)\left(1-G\left(\lambda;N\right)\right)}\right)\right\}=$$ $$=v-p^\ast\left\{1+H(\lambda)\right\}$$
Given that $\lim_{\lambda\to\hat{\lambda}}{\mathrm{\Gamma}\left(\lambda;N\right)}= v-v\left\{1+H(\hat{\lambda})\right\}<0$ and $\lim_{\lambda\to1}{\mathrm{\Gamma}\left(\lambda;N\right)}= v-s>0$, in order to show that equation (1) has a unique internal point of minimum it is sufficient to show that 
\begin{equation}
\frac{\partial\Gamma\left(\lambda;N\right)}{\partial\lambda}=-\left[\frac{\partial p^\ast}{\partial\lambda}\left\{1+H(\lambda)\right\}+p^\ast\frac{\partial H(\lambda)}{\partial\lambda}\right]>0    
\hspace{0.5cm} \forall \hspace{0.1cm} \lambda\in (\hat{\lambda},1)
\end{equation}

As $\frac{\partial p^\ast}{\partial\lambda}<0$  $\forall$ $\lambda\in (\hat{\lambda},1)$ and the reservation price is clearly positive, to show that equation (3) holds it is sufficient to show that $\frac{\partial H(\lambda)}{\partial\lambda}<0$ $\forall$ $\lambda \in (0,1)$. \\
 Algebraic manipulations reveal that $$\frac{\left(N-1\right)^2\left(1+\lambda\left(N-1\right)\right)^2\left(1-G\left(\lambda;N\right)\right)^2}{N^2}\cdot\frac{\partial H(\lambda)}{\partial\lambda}=\left(1-\frac{G\left(\lambda;N\right)}{\left(1-\lambda\right)}\right)$$
So that the condition $G\left(\lambda;N\right)\geq{1-\lambda}$ $\forall$ $\lambda \in (0,1)$ is sufficient to show that equation (3) indeed holds. 
\subsection{Solving the inequality}
\noindent
Our ultimate objective is, therefore, to show that:
\begin{equation}
G\left(\lambda;N\right)-\left(1-\lambda\right)\geq0\ \forall\ \lambda\ \in\ \left(0,1\right)\ ,\ \forall\ \ N>2\in\ \mathbb{N}
\end{equation}
\par
	To this end, it is necessary to exploit the Leibniz rule\footnote{See Mukhopadhyay(2000).} and the Bounded Convergence Theorem for the Riemann integral\footnote{See Gordon(2000).}. These two results allow us to state that $\lim_{\lambda\to0}{G^{\prime}\left(\lambda;N\right)=-1}$, $\lim_{\lambda\to0}{G^{\prime\prime}\left(\lambda;N\right)>0}$ and $\frac{\partial G^{\prime\prime}\left(\lambda;N\right)}{\partial\lambda}<0$ $\forall$ $\lambda\in (0,1)$. Importantly, it has been already established that $\lim_{\lambda\to0}{G\left(\lambda;N\right)=1}$ and $\lim_{\lambda\to1}{G\left(\lambda;N\right)=0}$\footnote{See Janssen et al (2005) in this regard.}.\\Let us show first that $\lim_{\lambda\to0}{G^{\prime}\left(\lambda;N\right)=-1}$, where, by the Leibniz rule, $$G^{\prime}\left(\lambda;N\right)=-\int_{0}^{1}{\frac{Ny^{N-1}}{{[(1+\lambda(Ny^{N-1}-1)]}^2}dy}$$ 
It is useful to write $$\lim_{\lambda\to0}{\int_{0}^{1}{\frac{Ny^{N-1}}{{[(1+\lambda(Ny^{N-1}-1)]}^2}dy}}=\lim_{n\to\infty}{\int_{0}^{1}{\frac{Ny^{N-1}}{{[(1+\lambda_{n}(Ny^{N-1}-1)]}^2}dy}}$$ where $\lbrace\lambda_{n}\rbrace$ is a generic sequence which satisfies $\forall$ $n \in \mathbb{N}$, $\lambda_{n} \in (0,1)$ and $\lim_{n\to\infty}{\lbrace\lambda_{n}\rbrace}=0$. Clearly, $\forall$ $n$ $$f_n(y)=\frac{Ny^{N-1}}{\left((1+\lambda_n(Ny^{N-1}-1)\right)^2}$$ is a Riemann-integrable function in $[0,1]$. Moreover, $\left\{f_n(y)\right\}$  converges pointwise to $f(y)=Ny^{N-1}$ in $[0,1]$, which is a Riemann-integrable function. Hence, to apply the Bounded Convergence Theorem, it remains to show that the sequence $\left\{f_n(y)\right\}$ is uniformly bounded. This motivates the following proposition. \\\\ \textit{Proposition 1. The sequence $\left\{f_n(y)\right\}$ is uniformly bounded.}\\\
\textit{Proof}. The proof works by contradiction. 
Suppose $\exists$ $n^{*}$ such that $\nexists$ $M \in \mathbb{R_+}$ which satisfies $\left|f_{n^*}(y)\right|=f_{n^*}(y)\leq{M}$,$\forall$ $y\in [0,1]$. Consider, for an arbitrary chosen $\varepsilon \in (0,1)$, the set $A_\varepsilon$=$\lbrace$$\lambda_n$ $\not\in$ $B_{\varepsilon}(0)$$\rbrace$, where $B_\varepsilon(0)$ is a neighbourhood of $0$ of radius $\varepsilon$. For any $\varepsilon$, the set $A_\varepsilon$ is either empty or finite. Suppose it is empty, so that $\lambda_{n^{*}} \in B_\varepsilon(0)$. Then, $\forall$ $y \in [0,{\frac{1}{N}}^{\frac{1}{N-1}}]$, $\forall$ $n$
  
 \begin{equation*}
f_\varepsilon(y)=\frac{Ny^{N-1}}{\left((1+\varepsilon(Ny^{N-1}-1)\right)^2}\geq{f_n(y)}
\end{equation*}
While $\forall$ $y \in [{\frac{1}{N}}^{\frac{1}{N-1}},1]$, $f(y)=Ny^{N-1}\geq{f_n(y)}$, $\forall$ $n$. \\
As a result, set 
$M=\max{\lbrace\max_{y \in [0,{\frac{1}{N}}^{\frac{1}{N-1}}]}{f_\varepsilon(y)},N\rbrace}\geq{f_{n^*}(y)}$ and the desired contradiction is reached. \\ Let $A_\varepsilon$ be finite. Then, $\exists$ $\lambda_m=\max{\lambda_n \in A_\varepsilon}$. Then, define $$K= \max_{y \in [0,{\frac{1}{N}}^{\frac{1}{N-1}}]}{f_{\lambda_m}(y)}$$
Setting $M=\max\lbrace{K,N}\rbrace$ we reach the desired contradiction, independently from whether $\lambda_{n^{*}} \in A_\varepsilon$ or not. The existence of the upper bound relies on the continuity of the functions $f_\varepsilon(y)$ and $f_{\lambda_m}(y)$, which are defined on a compact interval. \hfill $\blacksquare$
\\ \\
As a consequence, from the Bounded Convergence Theorem:$$\lim_{n\to\infty}{\int_{0}^{1}{\frac{Ny^{N-1}}{{[(1+\lambda_{n}(Ny^{N-1}-1)]}^2}dy}}=\int_{0}^{1}{Ny^{N-1}dy}=1$$ and  $\lim_{\lambda\to0}{G^{\prime}(\lambda;N)}=-1$. 
\\ Let us show the results for $G^{\prime\prime}(\lambda;N)$. Given 
$N$ and $y$, $f_{N,y}(\lambda)=\frac{Ny^{N-1}}{\left((1+\lambda(Ny^{N-1}-1)\right)^2}$ is differentiable for $\lambda$ in $(0,1)$. Thanks to this observation, it is licit to apply the Leibniz Rule, concluding that: $$G^{\prime\prime}(\lambda;N)= \int_{0}^{1}{\frac{2Ny^{N-1}(Ny^{N-1}-1)}{{(1+\lambda(Ny^{N-1}-1)}^3}dy}$$
To evaluate $\lim_{\lambda\to0}{G^{\prime\prime}(\lambda;N)}$, define the sequence, for a given $N$, $\lbrace g_{n} \rbrace$, with generic element $$ g_{n}(y)=\frac{2Ny^{N-1}(Ny^{N-1}-1)}{{(1+\lambda_{n}(Ny^{N-1}-1))}^3}$$ where $\lbrace\lambda_{n}\rbrace$ is a generic sequence which satisfies $\forall$ $n \in \mathbb{N}$, $\lambda_{n} \in (0,1)$ and $\lim_{n\to\infty}{\lbrace\lambda_{n}\rbrace}=0$. We have that $\lbrace g_{n} \rbrace$ is a sequence of Riemann integrable functions, since every $g_{n}$ is a continuous function defined on a compact interval. Moreover, the sequence converges pointwise to $g(y)=2Ny^{N-1}(Ny^{N-1}-1)$ in $[0,1]$, which is a Riemann Integrable function. Uniform boundedness\footnote{The proof is identical in spirit to the one of \textit{Proposition 1}.} of $\lbrace g_{n} \rbrace$ ensures that the Bounded Convergence Theorem holds, so that we are allowed to write: 
$$ \lim_{n\to\infty}{\int_{0}^{1}{\frac{2Ny^{N-1}(Ny^{N-1}-1)}{{(1+\lambda_{n}(Ny^{N-1}-1)})^3}dy}}=\int_{0}^{1}{2Ny^{N-1}\left(Ny^{N-1}-1\right)dy}=$$ $$=2N\left(\frac{N}{2(N-1)+1}-\frac{1}{N}\right)>0 \hspace{0.3cm} \forall \hspace{0.1cm} N>2$$ So that $\lim_{\lambda\to0}{G^{\prime\prime}(\lambda;N)}>0$. 
Finally, it is possible to notice that $G^{\prime\prime}(\lambda;N)$ is strictly decreasing in $\lambda \in (0,1)$. For every given $N$ and every given $y$, $f_{N,y}(\lambda)=\frac{2Ny^{N-1}(Ny^{N-1}-1)}{{(1+\lambda(Ny^{N-1}-1))}^3}$ is differentiable in $\lambda \in (0,1)$. By the Leibniz Rule, 
$$\frac{\partial}{\partial\lambda}\int_{0}^{1}{\frac{2Ny^{N-1}(Ny^{N-1}-1)}{{(1+\lambda(Ny^{N-1}-1))}^3}dy}=\int_{0}^{1}{-\frac{6Ny^{N-1}{(Ny^{N-1}-1)}^2}{{(1+\lambda(Ny^{N-1}-1))}^4}dy}$$ 

Given that $$\int_{0}^{1}{\frac{Ny^{N-1}{(Ny^{N-1}-1)}^2}{{(1+\lambda(Ny^{N-1}-1))}^4}dy>}\int_{0}^{1}{\frac{Ny^{N-1}{(Ny^{N-1}-1)}^2}{{(1+\lambda(N-1))}^4}dy}=$$ 
$$=\frac{1}{{(1+\lambda(N-1))}^4}\cdot\lbrace\frac{N^3}{3(N-1)+1}+1-\frac{{2N}^2}{2(N-1)+1}\rbrace>0 \hspace{0.2cm} \forall \hspace{0.1cm} N \in \mathbb{N}>2$$
Then $\frac{\partial}{\partial\lambda}G^{\prime\prime}(\lambda;N)<0$ $\forall$ $\lambda \in (0,1)$. This results is crucial in motivating the following proposition. \\\\
\textit{Proposition 2.}\\ $\forall$ $N \in \mathbb{N}>2$, \textit{there exists a unique} $\lambda_N^f\in (0,1)$ \textit{such that} $G^{\prime\prime}(\lambda_N^f;N)=0$.
\\
\textit{Proof}. Uniqueness stems directly from the strict and decreasing monotonicity of the function $G^{\prime\prime}(\lambda;N)$. Suppose that the inflection point did not exist. Then, $G^{\prime\prime}(\lambda;N)>0$ $\forall$ $\lambda \in (0,1)$ and, as a consequence, $G^{\prime}(\lambda;N)>-1$ $\forall$ $\lambda \in (0,1)$. Given that $\lim_{\lambda\to0}G(\lambda;N)=1$, this would contradict the result of $\lim_{\lambda\to1}G(\lambda;N)=0$.\hfill $\blacksquare$
\\
So far, we have showed that $\forall$ $N \in \mathbb{N}$, exists a unique $\lambda_N^f\in\ \left(0,1\right)$ such that $G^{\prime}\left(\lambda;N\right)>-1$ $\forall$ $\lambda\in (0,\lambda_N^f]$ and the function $G\left(\lambda;N\right)$ is strictly concave in $(\lambda_N^f,1)$. This preliminary conditions allow us to show the desired inequality. \\\\
\textit{Proposition 3}. 
$$G(\lambda;N)-(1-\lambda)\geq0 \hspace{0.2cm} \forall\hspace{0.1cm} \lambda \in\ (0,1), \hspace{0.2cm} \forall\hspace{0.1cm} N>2 \in \mathbb{N}$$
\textit{Proof}. Suppose $\exists$ $\bar{\lambda} \in (0,\lambda_N^f]$ such that $G(\bar{\lambda};N)<(1-\bar{\lambda})$; this would contradict the results of $\lim_{\lambda\to0}{G(\lambda;N)}=1$ and $\forall$ $\lambda \in (0,\lambda_N^f] \ G^{\prime}(\lambda;N)>-1$. Suppose $\exists$ $\hat{\lambda} \in (\lambda_N^f,1)$ such that $G(\hat{\lambda};N)<(1-\hat{\lambda})$. Then, given that $G(\lambda_N^f;N)>(1-\lambda_N^f)$ and $G^{\prime\prime}(\lambda;N)<0$ $\forall$ $\lambda \in (\lambda_N^f,1)$, the condition $\lim_{\lambda\to1}{G(\lambda;N)}=0$ would be impossible to be satisfied. \hfill $\blacksquare$\\
Thanks to \textit{Proposition 3}, we are allowed to state that, given $N \in \mathbb{N}$ and $s \in (0,v)$, $\exists$ $\widetilde{\lambda}$ such that $\forall\ \lambda\in\ (\hat{\lambda},\widetilde{\lambda}]$ $\frac{\ \partial\delta^\ast}{\partial\lambda}\ \leq0$ and $\forall\ \lambda\in\ \left(\widetilde{\lambda},1\right)$ $\frac{\ \partial\delta^\ast}{\partial\lambda}>0$.
\section*{References}
\noindent
\textbf{Gordon, R. (2000)}: \textit{A Convergence Theorem for the Riemann Integral}. Mathematics Magazine, 73(2), 141-147.\\\textbf{Janssen, M., Moraga-González, J.L. and Wildenbeest, M. (2005)}: \textit{Truly Costly Sequential Search and Oligopolistic Pricing}, International Journal of Industrial Organization 23, 451-466. \\ \textbf{Motta, M. (2004)}: \textit{Competition Policy: Theory and Practice}, Cambridge University Press.\\ \textbf{Mukhopadhyay, N.  (2000)}: \textit{Probability and Statistical Inference}, New York: Marcel Dekker. \\\textbf{Petrikaite, V. (2016)}: \textit{Collusion with Costly Consumer Search}, International Journal of Industrial Organization vol. 44, pp. 1-10.
\newpage
\section*{List of Figures}
\begin{figure}[!h]
\includegraphics[width=\textwidth]{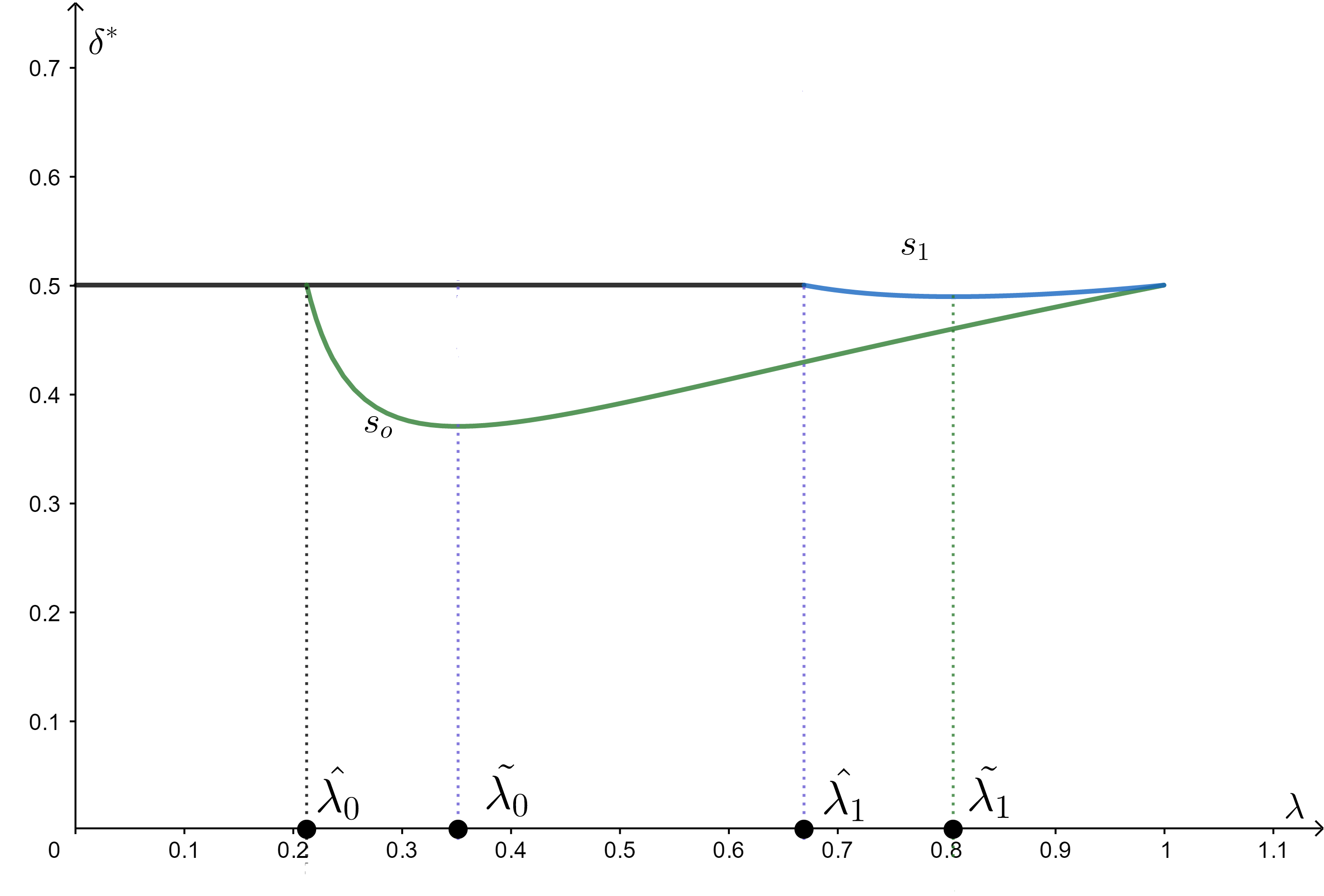}
\caption{Critical discount factor as a function of the proportion of shoppers in a duopoly. Here, $v=1$, $s_0=0.2$, $s_1=0.6$.}
\end{figure}
\newpage
\begin{figure}[!h]
\includegraphics[width=\textwidth]{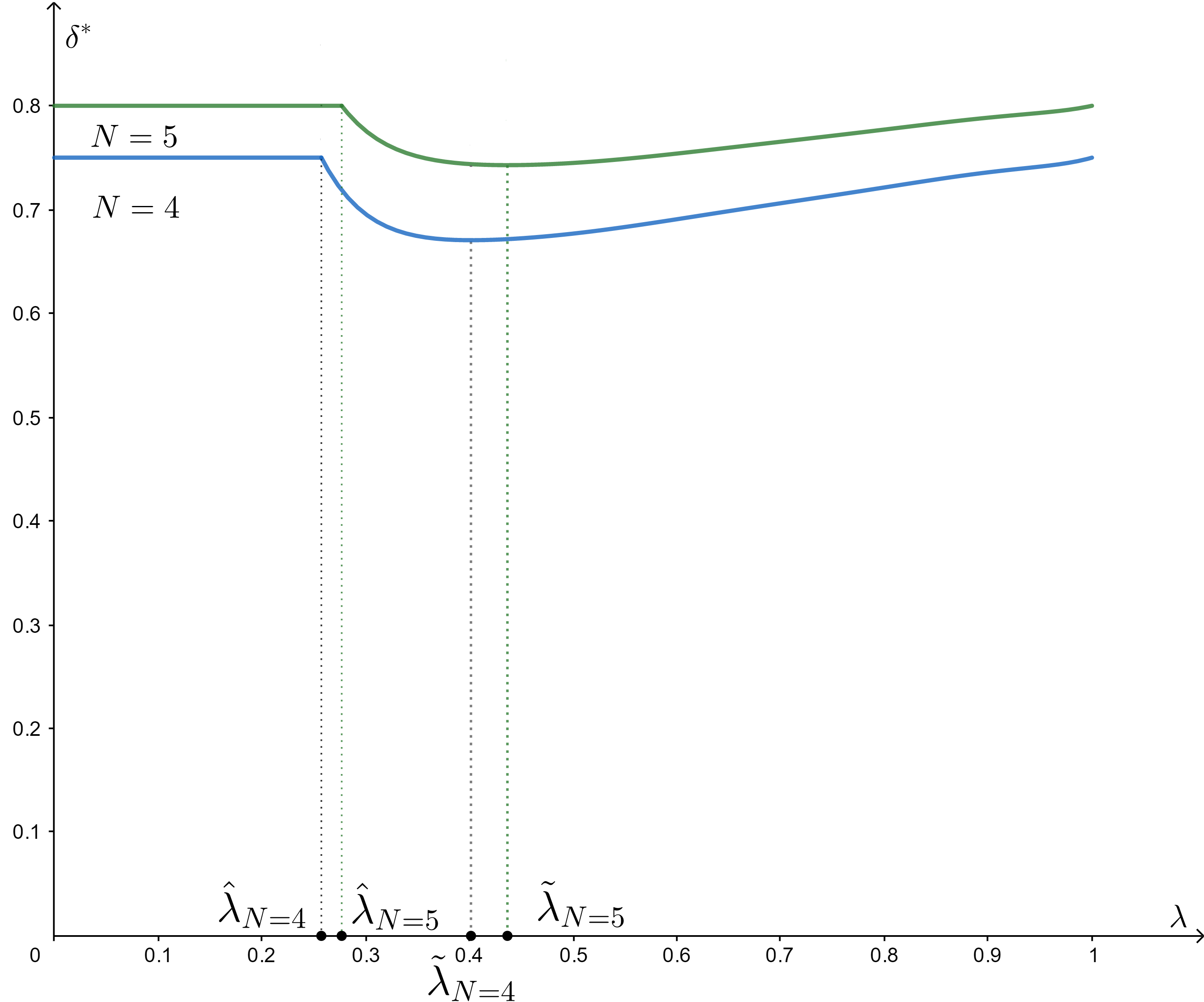}
\caption{Critical discount factor as a function of the proportion of shoppers when $N=4$ and $N=5$. Here, $v=1$ and $s=0.2$ fixed. Numerical approximations show that the duopoly pattern is confirmed also when a higher number of firms interacts.}
\end{figure}
\newpage
\begin{figure}[!h]
\includegraphics[width=\textwidth]{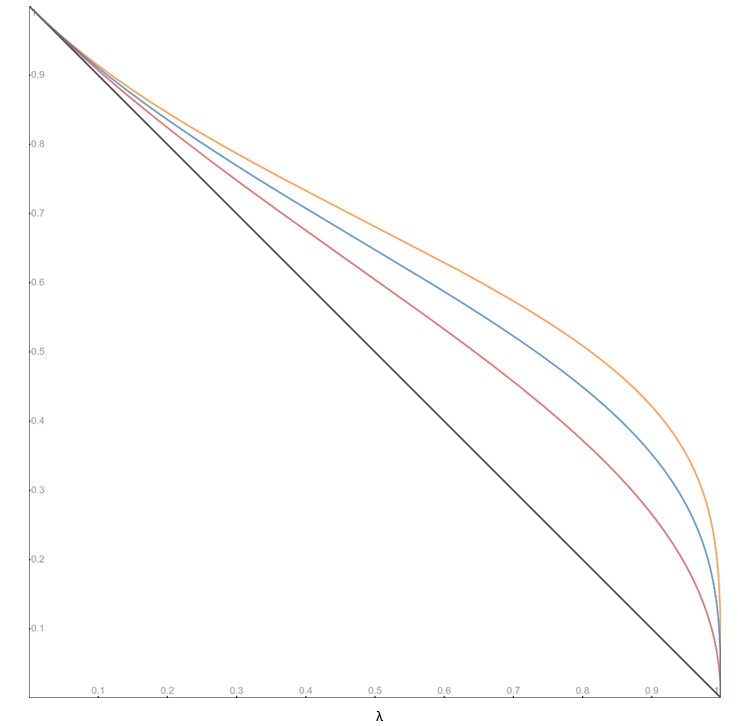}
\caption{Numerical approximations confirm the desired inequality. The black line indicates $f(\lambda)=1-\lambda\leq {G(\lambda;3)}\leq{G(\lambda;4)}\leq{G(\lambda;5)}$,where $\lambda\in (0,1)$.} 
\end{figure}

\end{document}